\documentclass[aps,pre,twocolumn,groupedaddress, amsmath, 10pt, floatfix, longbibliography]{revtex4-2}
\usepackage{bm}
\usepackage{graphicx}
\usepackage[caption=false]{subfig}
\usepackage[normalem]{ulem}
\usepackage{xcolor}

\newcommand{\Add}[1]{{#1}}
\newcommand{\Delete}[1]{}

\begin{document}

\title{Defect binding-unbinding transition in active nematic membranes
}

\author{Yuki Hirota}
\author{Nariya Uchida}
\email{nariya.uchida@tohoku.ac.jp}
\affiliation{Department of Physics, Tohoku University, Sendai 980-8578, Japan}

\date{\today}

\begin{abstract}
We investigate the dynamics of active nematic liquid crystals on deformable membranes, focusing on the interplay between active stress and anisotropic curvature coupling. 
Using a minimal model, we simulate the coupled evolution of the nematic order parameter and membrane height. We demonstrate a continuous transition from a curvature-dominated regime, where topological defects are trapped by local deformation, 
to an activity-dominated regime exhibiting active turbulence. 
A scaling analysis reveals that the critical activity threshold 
$\zeta_c$ scales as $\alpha^2/\kappa$, where $\alpha$ and $\kappa$ are the coupling constant and bending stiffness, respectively; this relationship is 
confirmed by our numerical results. 
Furthermore, we find that significant correlations between the orientational pattern and membrane geometry persist even in the turbulent regime. Specifically, we identify that "walls" in the director field induce characteristic wave-like curvature profiles, providing a mechanism for
dynamic coupling between order and shape. 
These results offer a physical framework for understanding defect-mediated deformation in nonequilibrium biological membranes.
\end{abstract}

\maketitle

\section{Introduction}

Biological systems such as the cytoskeleton and epithelial tissues are representative examples of active matter, in which constituent elements consume energy to generate mechanical forces~\cite{marchetti2013hydrodynamics, ramaswamy2010mechanics, bechinger2016active, bowick2022symmetry}. 
In particular, systems containing anisotropic components, such as actin filaments, microtubules, or cells themselves, that exhibit orientational order are modeled as active nematics~\cite{doostmohammadi2018active}. 
This concept provides a unified framework for understanding biological phenomena ranging from intracellular cytoplasmic streaming to the collective motion of multicellular tissues~\cite{balasubramaniam2022active, prost2015active}. 

The most prominent feature of active nematics is a spatiotemporal chaotic state known as active turbulence, characterized by the continuous creation and annihilation of topological defects accompanied by spontaneous flows~\cite{sanchez2012spontaneous, wensink2012meso, lemma2019statistical, giomi2015geometry, keber2014topology, ellis2018curvature}. 
At first glance, this turbulent state appears antithetical to biological morphogenesis. However, recent studies have revealed that these topological defects are not merely disorders but play crucial roles in regulating biological functions. 
For instance, defects trigger apoptosis and cell extrusion in epithelial monolayers~\cite{saw2017topological}, and serve as nucleation sites for cell accumulation in neural progenitor cells~\cite{kawaguchi2017topological}. 
In particular, in {\it Hydra} regeneration, defects have been shown to function as organizational centers for specific morphological features such as the mouth and foot~\cite{maroudas2021topological}. In {\it Hydra}, actin filaments in the outer layer (ectoderm fibers) run parallel to the body axis, which suggests the physical importance of an anisotropic coupling between the orientation and curvature. Furthermore, recent research reports the concentration of mechanical strain at defect locations~\cite{maroudas2025mechanical}, suggesting a strong coupling between the orientational defect structures and the geometric shape of the membrane. 

Biological tissues and cell membranes, which serve as the arena for such active behaviors, possess intriguing mechanical properties. 
While tissues exhibit elastic responses on short timescales, they behave macroscopically as viscous fluids on longer timescales associated with development and regeneration, driven by cell rearrangement and remodeling~\cite{wang2020anisotropy, petridou2019fluidization, chen2022self}. 
Indeed, many biological interfaces, such as developing epithelial tissues, display fluid-like characteristics, where shape deformation is intimately coupled with internal flow~\cite{etournay2016tissueminer, erdemci2021effect}. 
This nature of a flowing, deformable membrane is an indispensable element for the physical understanding of biological morphogenesis.

Previous research has explored active matter dynamics on deformable membranes or coupled to geometry\Add{~\cite{mietke2019self, bacher2019computational, vafa2022active, al2023morphodynamics, al2021active, salbreux2022theory}}.
Notable examples include frameworks linking curvature dynamics to defects and cell growth~\cite{vafa2022active}, instability analyses of active nematic surfaces~\cite{al2023morphodynamics}, 
and defect-mediated morphogenesis driven by polar order~\cite{hoffmann2022theory} or on fixed curved manifolds~\cite{wang2023patterning}. 
\Add{Recent developments have further advanced the nematohydrodynamics on deformable surfaces using various theoretical and numerical approaches~\cite{nitschke2025active, mirza2025variational, zhu2025active}. Additionally, the interplay between topology and morphology has been investigated in self-deforming active shells~\cite{metselaar2019topology} and patterned epithelial layers~\cite{khoromskaia2023active}.}
Despite these advances, the role of anisotropic coupling between the orientational degrees of freedom and the membrane curvature remains a critical yet unexplored frontier in the physics of active nematic membranes. In such systems, the energetic significance of the director's alignment relative to the principal curvatures~\cite{uchida2002dynamics}
is expected to fundamentally alter the active defect dynamics.

In this paper, we simulate a minimal model of an active nematic membrane with an anisotropic coupling, extending the Ginzburg-Landau framework established  for passive systems~\cite{uchida2002dynamics}  to the active regime. 
Using this model, we demonstrate 
an activity-induced transition of defect structures
from a curvature-dominated regime to an activity-dominated regime. 
We show that the critical threshold for this transition obeys a specific scaling law, $\zeta_c \propto \alpha^2 / \kappa$.
Furthermore, we find that the activity-dominated regime is characterized by significant correlations between the membrane geometry and director walls~\cite{thampi2014instabilities}. Specifically, these walls induce wave-like curvature profiles, providing a mechanism for dynamic coupling between order and shape in active surfaces.

\section{Model}
\subsection{Free energy and governing equations}

We describe the nematic order of a two-dimensional liquid crystal on a membrane using a symmetric and traceless tensor field $\bm{Q}$. 
In two dimensions, this order parameter tensor is represented as  $Q_{ij} = S ( n_i n_j - \delta_{ij} / 2 )$ $(i, j = x, y)$,
where $S$ is the scalar order parameter denoting 
the magnitude of the ordering, and $\bm{n} = (\cos\theta, \sin\theta)$ is the director field representing the local average orientation. 
We consider a nearly flat fluid membrane in the Monge gauge, where its shape
is described by the height $z = h(x, y)$ 
satisfying the small-gradient approximation $|\nabla h| \ll 1$\Add{~\cite{napoli2012surface}}. 

The total free energy $F_{\text{total}}$ consists of the Landau-de Gennes free energy $f_{\text{LdG}}$, the bending energy $f_{\text{bend}}$ and the coupling energy $f_{\text{couple}}$ :
\begin{align}
    F_{\text{total}} &= \int \left( f_{\text{LdG}} + f_{\text{bend}} + f_{\text{couple}} \right) d\bm{r}, \label{eq:F_total} \\
    f_{\text{LdG}} &= \frac{A}{2} \text{Tr}(\bm{Q}^2) + \frac{B}{4} (\text{Tr}(\bm{Q}^2))^2 + \frac{M}{2} (\nabla \bm{Q})^2, \label{eq:f_LdG} \\
    f_{\text{bend}} &= \frac{\kappa}{2} (\nabla^2 h)^2, \label{eq:f_b}\\
    f_{\text{couple}} &= \alpha \bm{Q} : \nabla \nabla h, 
    \label{eq:f_c_2}
\end{align}
Here, $A$, $B$, and $M$ are Landau-de Gennes coefficients, $\kappa$ is the bending rigidity, $\alpha$ is the coupling constant. 
In Eq.~\eqref{eq:f_LdG}, the cubic term proportional to $\text{Tr}(\bm{Q}^3)$ vanishes for a two-dimensional order parameter, 
and we adopt the one-constant approximation
for the Frank elastic energy density. 
Equation~\eqref{eq:f_c_2} represents the coupling term in a nonchiral nematic membrane~\cite{uchida2002dynamics}. 

The dynamics of the active nematic system are governed by the coupled equations for the fluid velocity field $\bm{v}$ and the nematic order parameter $\bm{Q}$~\cite{doostmohammadi2018active}. 
At the low Reynolds number limit where inertial effects are negligible, the velocity field satisifies the Stokes equation:
\begin{equation}
    \eta \nabla^2 \bm{v} - \nabla p + \nabla \cdot \bm{\Sigma} = 0, 
    \label{eq:v}
\end{equation} 
coupled with the incompressibility condition $\nabla \cdot \bm{v} = 0$. Here, $p$ denotes the pressure, $\eta$ is the fluid viscosity, and $\bm{\Sigma}$ is the stress tensor.
The temporal evolution of the tensor field $\bm{Q}$ obeys the following equation: 
\begin{equation}    
    \frac{D \bm{Q}}{Dt} + \bm{\omega} \cdot \bm{Q} - \bm{Q} \cdot \bm{\omega} = \lambda S \bm{u} + \Gamma_Q \bm{H}, 
    \label{eq:Q}
\end{equation}
where $D / Dt = \partial / \partial t + \bm{v} \cdot \nabla$ is the material derivative\Add{~\cite{pollard2025gauge}}, 
$\lambda$ is the flow-aligning parameter and $\Gamma_Q$ is the reciprocal of the rotational viscosity.
The tensors $u_{ij} = (\partial_i v_j + \partial_j v_i) / 2$ and $\omega_{ij} = (\partial_i v_j - \partial_j v_i) / 2$ are the symmetric and antisymmetric parts of the velocity gradient tensor, respectively.
The molecular field $H_{ij}$, which drives the relaxation of the order parameter, is obtained from $- (\delta F_{\text{total}} / \delta \bm{Q})^{(s)}$ ($(s)$ denotes the symmetric and traceless part). This yields:
\begin{equation}
    \bm{H} = -(A + B \text{Tr}(\bm{Q}^2)) \bm{Q} + M \nabla^2 \bm{Q} - \alpha \bm{C}.
\end{equation}
Here, $\bm{C}$ is the traceless part of the curvature tensor (the Hessian of $h$), defined as $C_{ij} = \partial_i \partial_j h - (\delta_{ij} / 2) \nabla^2 h$. Since $\bm{Q}$ is traceless, the coupling energy $\alpha \bm{Q} : \nabla \nabla h$ is identically equal to $\alpha \bm{Q} : \bm{C}$.
The stress tensor is given by the sum of the passive stress $\bm{\Sigma}^{(\text{p})}$ and the active stress $\bm{\Sigma}^{(\text{a})}$:  
\begin{align}
    \bm{\Sigma} &= \bm{\Sigma}^{(\text{p})} + \bm{\Sigma}^{(\text{a})}, \\
    \bm{\Sigma}^{(\text{p})} &= -\lambda S \bm{H} + \bm{Q} \cdot \bm{H} - \bm{H} \cdot \bm{Q}, \\
    \bm{\Sigma}^{(\text{a})} &= - \zeta \bm{Q}.
\end{align}
The parameter $\zeta > 0$ represents an extensile active stress.

Finally, the dynamical equation for the height is 
assumed to follow a purely relaxational form:
\begin{equation}
    \frac{\partial h}{\partial t} = - \Gamma_h \frac{\delta F_{\text{total}}}{\delta h} = -\Gamma_h (\kappa \nabla^2 \nabla^2 h + \alpha \nabla \nabla : \bm{Q}).
    \label{eq:h}   
\end{equation}
This formulation is based on the observation that a constant fluid flow translates membrane components without altering the height profile, provided the membrane is flat or has a constant slope $\nabla h$. Due to incompressibility and this kinematic constraint, the leading-order scalar coupling between the velocity and the height gradient must be proportional to $\nabla^2\bm{v} \cdot \nabla h$. In the small-gradient limit considered here, such a term is neglected as a higher-order contribution.
\Add{The effects of complex couplings between the fluid flow and the intrinsic curvature of 
the membranes have been explored in previous studies on curved fluid interfaces~\cite{arroyo2009relaxation, henle2010hydrodynamics, morris2015mobility}.}

\subsection{Anisotropic curvature coupling}

The coupling energy defined in the previous section induces a correlation between the director field and the principal directions of the curvature. 
As noted above, the curvature contribution is captured by the traceless tensor $\bm{C} = \nabla \nabla h - (\nabla^2 h / 2) \bm{I}$.
Since the eigenvectors of $\bm{C}$ correspond to the principal axes of the local curvature, this energy term implies that the director tends to align along these geometric axes.

Explicitly, the coupling energy is proportional to $\alpha \cos (2(\theta - \phi))$, where $\theta$ is the director angle and $\phi$ is the angle of the eigenvector associated with the positive eigenvalue of $\bm{C}$.
For $\alpha > 0$, the energy is minimized when the director aligns with the direction of minimum curvature (i.e., perpendicular to the direction of maximum bending). 
In contrast, for $\alpha < 0$, the director prefers to align with the direction of maximum curvature. 
Therefore, when the coupling energy is dominant, it is expected that a strong correlationemerges between the anisotropic  curvature  $\bm{C}$ and the order parameter tensor $\bm{Q}$ in the steady state.    

\subsection{Local equilibrium condition}

We first consider the local mechanical equilibrium condition, $\delta F_{\text{total}} / \delta h = 0$, following the framework established for passive systems. Under this condition, Eq.~\eqref{eq:h} leads to:
\begin{equation}
\nabla^2 \nabla^2 h = - \frac{\alpha}{\kappa} J(\bm{r}),
\label{eq:steady_state}
\end{equation}
where the source term is defined as:\begin{equation}J(\bm{r}) = \nabla \nabla : \bm{Q}.\label{eq:J}\end{equation}By performing a Fourier transform, we obtain the relationship 
$h(\bm{q}) = (\alpha / \kappa q^4) J(\bm{q})$. 
Substituting this into the bending and coupling energy terms, Eqs.~\eqref{eq:f_b} and~\eqref{eq:f_c_2}, 
we derive the effective curvature elastic energy:
\begin{align}
F_{\text{curv}} &= \Add{\int d \bm{r} \left( f_{\text{bend}} + f_{\text{couple}}\right) } \\
 &= - \frac{\alpha^2}{2 \kappa} \int \frac{d \bm{q}}{(2 \pi)^2} \frac{|J(\bm{q})|^2}{q^4}.\label{eq:e_curv}
\end{align}
This expression indicates that $F_{\text{curv}}$ is always non-positive and decreases as the magnitude of the source term increases. Crucially, the $q^{-4}$ factor implies that low-wavenumber components of $J(\bm{q})$ contribute predominantly to the energy reduction. This long-range interaction suggests that large-scale nematic textures are more effectively stabilized by membrane deformations than localized fluctuations.

\begin{figure}[h]
    \centering
    \includegraphics[width=\linewidth]{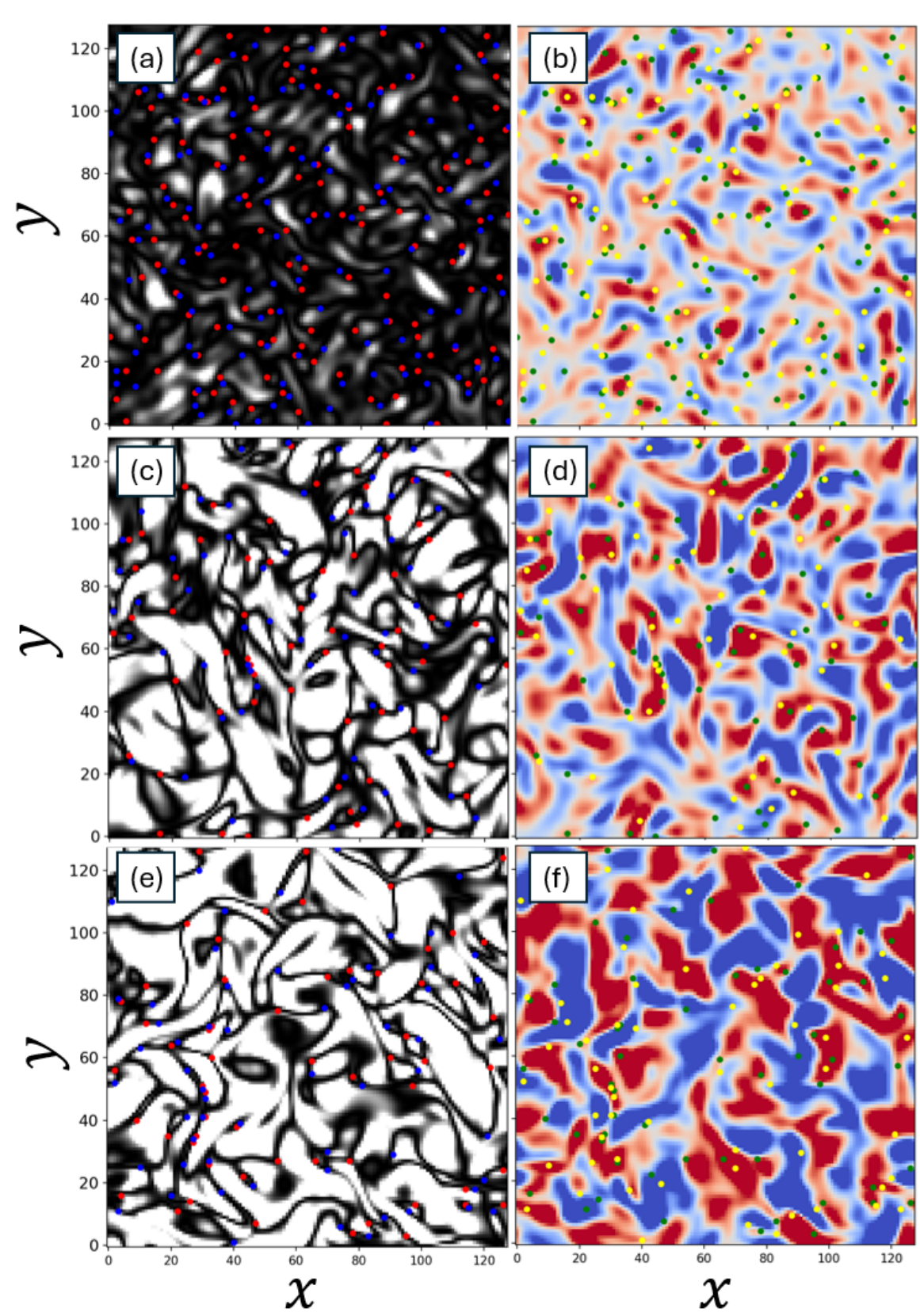}
    \caption{Snapshots of the Schlieren texture $Q_{xy}^2(\bm{r})$ (left column) and the height field $h(\bm{r})$ (right column) at times $t = 50, 100,$ and $400$. The parameters are set to $(\alpha, \zeta) = (10, 1.0)$. (a,c,e) The color map indicates the magnitude of $Q_{xy}^2$, where red and blue dots denote $+1/2$ and $-1/2$ defects, respectively. (b,d,f) The height field, where red and blue regions represent positive and negative values,  respectively. In these panels, the $+1/2$ and $-1/2$ defects are marked by green and yellow dots, respectively. 
    }
    \label{fig:pattern} 
\end{figure}

\section{Numerical results}
\subsection{Numerical methods and parameters}

We numerically solve the coupled dynamical equations for the order parameter $\bm{Q}$ (Eq.~\eqref{eq:Q}) and the membrane height $h$ (Eq.~\eqref{eq:h}) on a square lattice. 
The system size is $128 \times 128$ lattice points with a grid spacing of $\Delta x = \Delta y = 1$. We impose periodic boundary conditions for all fields. 
To solve the system, we first discretize the evolution equations using the finite difference method and then integrate them in time using the fourth-order Runge-Kutta method with a time step of $\Delta t = 0.01$. 
At each time step, the incompressible Stokes equation (Eq.~\eqref{eq:v}) is solved in the Fourier space to update the velocity field $\bm{v}$.
As the initial condition, the membrane is set to be flat ($h = 0$) and the components of the order parameter $\bm{Q}$ are distributed uniformly at random in the range $[-0.1, 0.1]$. 

In our simulations, the following fixed parameters are used: $A = -0.5$, $B = 2.0$, $M = 1.0$, $\lambda = 0.1$, $\Gamma_Q = 0.1$, $\Gamma_h = 0.01$, $\kappa = 100$.
To investigate the interplay between activity and curvature,  
we vary the coupling constant $\alpha$ (from 0 to 10), and the activity $\zeta$ (from 0 to 1.0). 
\Add{The physical scales of the system are determined by these parameters. 
We identify $l_{\rm core} = \sqrt{M/|A|} = \sqrt{2}$ as the characteristic length scale and $\tau_{\rm core} = l_{\rm core}^2/\Gamma_Q = 20$ as the characteristic time scale of the orientational relaxation. 
Under this scaling, the crossover length between the membrane and orientational relaxation is $l_c = \sqrt{\Gamma_h/\Gamma_Q} \approx 0.32$. 
Since the grid spacing $\Delta x = 1$ is significantly larger than $l_c$, the height relaxes slower than the orientation for all numerically resolved scales. 
For a typical correlation length $\xi_H \sim 7$ in the late stage, the relaxation time for the height profile is estimated as $\tau_{H} \sim \xi_H^2/(\Gamma_h \alpha S)\approx 50 \tau_{\rm core}$, confirming that membrane deformation is the rate-limiting process for the coarsening dynamics.}
Throughout the simulations, we verified that the spatial average of the height gradient satisfies
$\langle |\nabla h| \rangle < 0.2$,
ensuring that the nearly flat membrane approximation ($|\nabla h| \ll 1$) remains valid. \Add{To verify the local validity of the Monge gauge approximation ($\vert \nabla h \vert \ll 1$), we computed the distribution of the gradient magnitude $\vert \nabla h \vert$ over the late stage ($t \geq 2000$). For zero activity ($\zeta = 0.0$), the average is $\langle \vert \nabla h \vert \rangle \approx 0.14$, while the maximum absolute value is approximately 0.40. Because the relevant metric tensor components scale with $\left( \nabla h \right)^2$, the maximum local deviation from the flat plane approximation is about 16\%. In the high-activity regime ($\zeta = 1.0$), the maximum value is strongly suppressed to roughly 0.06. Thus, the small-gradient approximation holds robustly throughout the system without qualitative breakdown.}

\subsection{Spatiotemporal evolution of patterns}

In Fig.~\ref{fig:pattern}, we show representative snapshots of the Schlieren texture $Q^2_{xy} (\bm{r})$ (left column) 
and the membrane height $h(\bm{r})$ (right column)
during the time evolution from $t = 0$ to $t = 400$. 
As the system evolves, both the orientation field and the height field undergo coarsening, leading to the formation of 
ordered domains. 
In the later stage, the system reaches a statistically steady state of active turbulence, which is characterized by the continuous creation and annihilation of $+1/2$ and $-1/2$ defect pairs accompanied by vortex formation.  

\begin{figure*}[t]
    \centering
    \includegraphics[width=\linewidth]{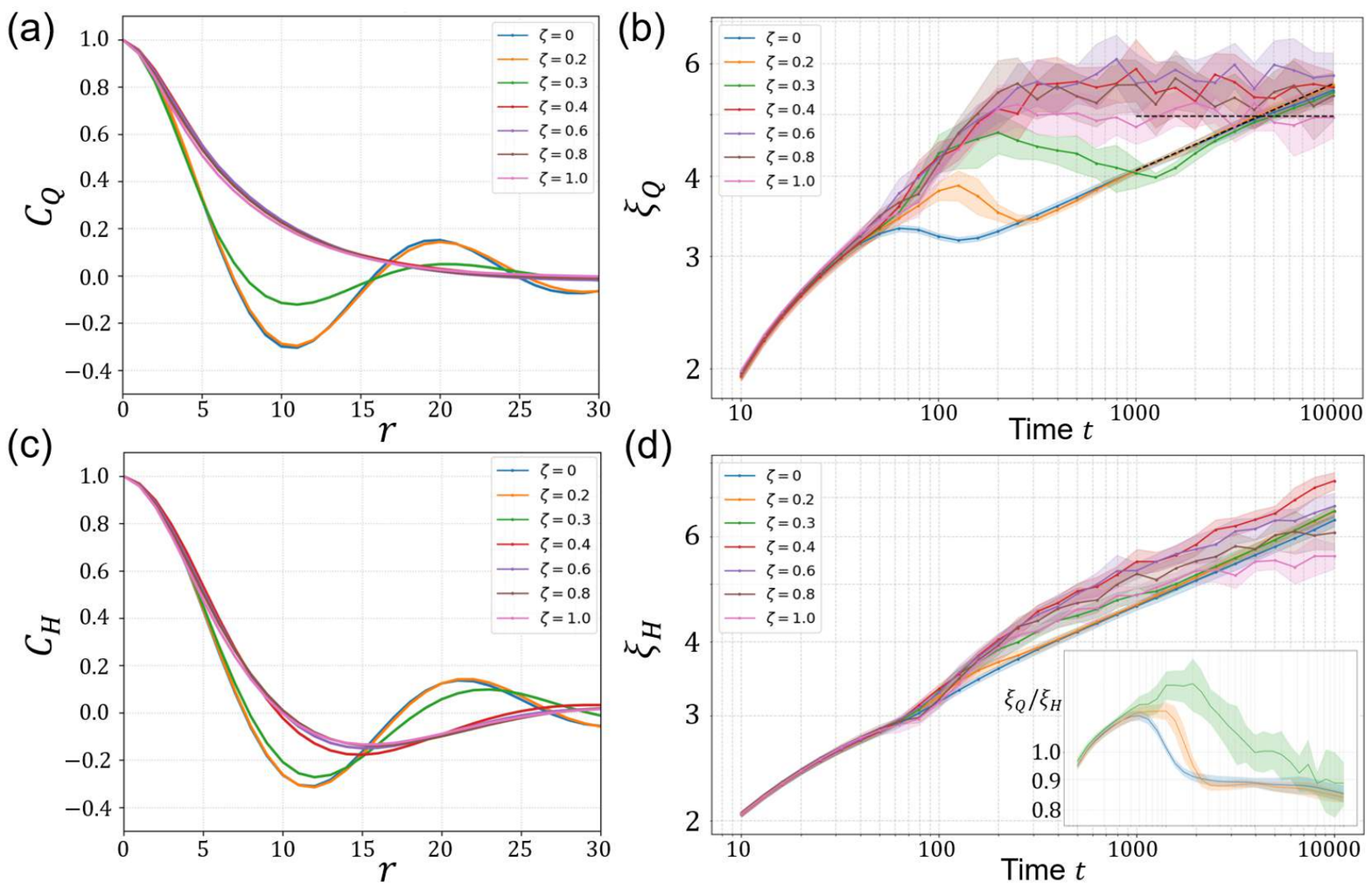}
    \caption{\Add{Spatial correlation functions and time evolution of correlation lengths. (a) Orientational correlation function $C_Q(r)$ and (c) height correlation function $C_H(r)$ at $t=1000$.} Time evolution of (b) the orientational correlation length $\xi_Q(t)$ and (d) the height correlation length $\xi_H(t)$ (log-log plot). The parameter $\alpha$ is fixed at $10$, while $\zeta$ is varied from $0$ to $1.0$ in increments of $0.2$. The data are averaged over 10 samples, and the colored regions indicate the standard deviation. In \Add{(b)}, the black dotted lines represent fitting results: a constant fit for $\zeta = 1.0$ (value $5.0$) and a power-law fit for $\zeta = 0.2$ (exponent $0.14$). \Add{(d)} The inset shows the ratio $\xi_Q / \xi_H$. \Add{The green, orange, and blue lines correspond to $\zeta = 1.0$, $0.2$, and $0$, respectively.}}
    \label{fig:corlength} 
\end{figure*}

To investigate this domain growth in more detail, we calculate the spatial correlation functions $C_Q(|\bm{r}-\bm{r}'|, t) = \langle Q_{ij}(\bm{r}, t) Q_{ij}(\bm{r}', t) \rangle - \langle Q_{ij}^2 \rangle$ 
and $C_H(|\bm{r}-\bm{r}'|, t) = \langle h(\bm{r}, t) h(\bm{r}', t) \rangle - \langle h^2 \rangle$.
\Add{Figures 2(a) and 2(c) show the full spatial correlation functions $C_Q(r)$ and $C_H(r)$ in the late stage ($t=1000$). In the low-activity regime ($\zeta \leq 0.3$), both correlation functions exhibit pronounced oscillatory behavior. The oscillation in $C_H(r)$ reflects the formation of a bumpy, highly deformed membrane structure. Furthermore, the corresponding oscillation in $C_Q(r)$ indicates that the orientational field is strongly dragged by this geometric deformation due to the anisotropic curvature coupling. 
Conversely, in the high-activity regime ($\zeta \geq 0.4$), these oscillations are strongly suppressed.}
The correlation lengths $\xi_Q(t)$ and $\xi_H(t)$ are then determined from these functions as the distance $r$ at which they decay to half of their initial value, i.e., $C(r=\xi) = C(r=0)/2$. 
We show the time evolution of these correlation lengths for different parameter sets in Fig.~\ref{fig:corlength}. The orientational correlation length, $\xi_Q(t)$ (Fig.\Add{~2(b)}), exhibits faster growth compared to the height correlation length, $\xi_H(t)$ (Fig.\Add{~2(d)}), prior to reaching the steady state. 
However, in the low activity regime ($\zeta \leq 0.3$), $\xi_Q(t)$ initially exhibits rapid growth comparable to the high-activity case, but subsequently merges onto the $\xi_H(t)$ curve. 
As shown in the inset of Fig.~2(b), the ratio of the correlation lengths, $\xi_Q / \xi_H$, converges to approximately $0.9$ in the late stage. 
This indicates that the defect coarsening is limited by the slow membrane deformation 
and extends the previous result for the passive case~\cite{uchida2002dynamics} 
to the low activity regime.

Based on the equation of motion Eq.~\eqref{eq:h}, the relaxation of the membrane height is governed by the fourth-order derivative term ($\nabla^2 \nabla^2 h$). 
This implies that the membrane evolves according to subdiffusive dynamics, resulting in a relaxation process that is significantly slower than the diffusive relaxation of the orientational field. Our fitting analysis for the case of $\zeta=0.2$ yields a growth exponent of approximately 0.14 in the low activity regime (black dotted line in Fig. 2(a)). 
This value is in good agreement with the slow coarsening dynamics reported in previous numerical simulations of passive fluid membranes, where the coupling to membrane deformation limits the texture growth\cite{uchida2002dynamics}. 
In contrast, at higher activity levels, the active stress dominates over the membrane relaxation. Consequently, the orientational correlation length $\xi_Q(t)$ decouples from the height correlation length
$\xi_H(t)$ and exhibits rapid growth, 
eventually saturating as the system enters an active turbulence. 

\subsection{Activity-induced transition}

\begin{figure}[h]
    \centering
    \includegraphics[width=\linewidth]{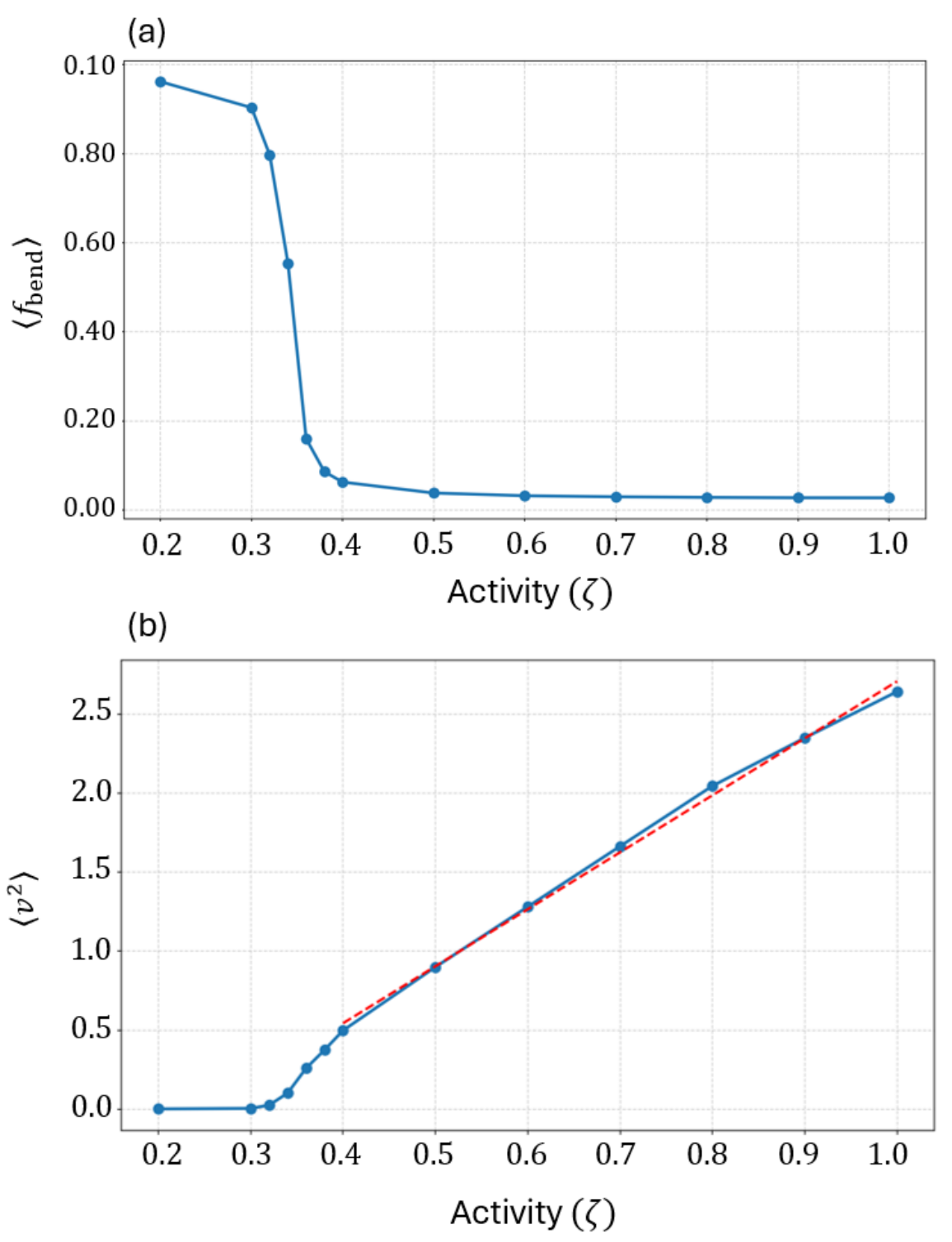}
    \caption{Activity dependence of the bending energy and fluid velocity. Each data point is time-averaged over the interval $t = 1000$ to $10000$. (a) The spatially averaged bending energy $\langle f_{\text{bend}} \rangle$ as a function of activity $\zeta$ for various $\alpha$. (b) Mean squared velocity $\langle v^2 \rangle$ for the same parameters. The red dashed line represents the linear fit $\langle v^2 \rangle = a (\zeta - 0.4) + b$ for $\alpha=10$, where $a = 3.61$, and $b = 0.544$.}
    \label{fig3:transition}
\end{figure}

The distinct growth laws observed in the coarsening dynamics suggest a fundamental change in the system's behavior driven by activity. 
To clarify the nature of the transition from the membrane-dominated regime to the active turbulent regime, we examine 
the time-averaged properties in the late stage
properties as a function of activity $\zeta$. 
In Fig.~\ref{fig3:transition}, we present the energetic and kinetic properties averaged over the time interval from $t=1000$ to $10000$  (for $\alpha=10$).
Fig.~\ref{fig3:transition}(a) presents the mean bending energy density, $\langle f_{\text{bend}} \rangle$. At a critical activity $\zeta_c \approx 0.35$, $\langle f_{\text{bend}} \rangle$ drops abruptly from a finite value to nearly zero, marking the onset of the transition. This suggests that the slow membrane deformation can no longer follow or sustain the ordered structures formed by active defects. Conversely, Fig.~\ref{fig3:transition}(b) plots the mean squared fluid velocity $\langle v^2 \rangle$ as a function of activity. The data show that $\langle v^2 \rangle$ 
remains negligible in the low-activity regime and
increases continuously once the activity exceeds
the critical point. Beyond this point, $\langle v^2 \rangle$ 
exhibits a linear increase with $\zeta$,
consistent with the behavior observed in conventional active turbulent systems.

\begin{figure}[h]
    \centering
    \includegraphics[width=\linewidth]{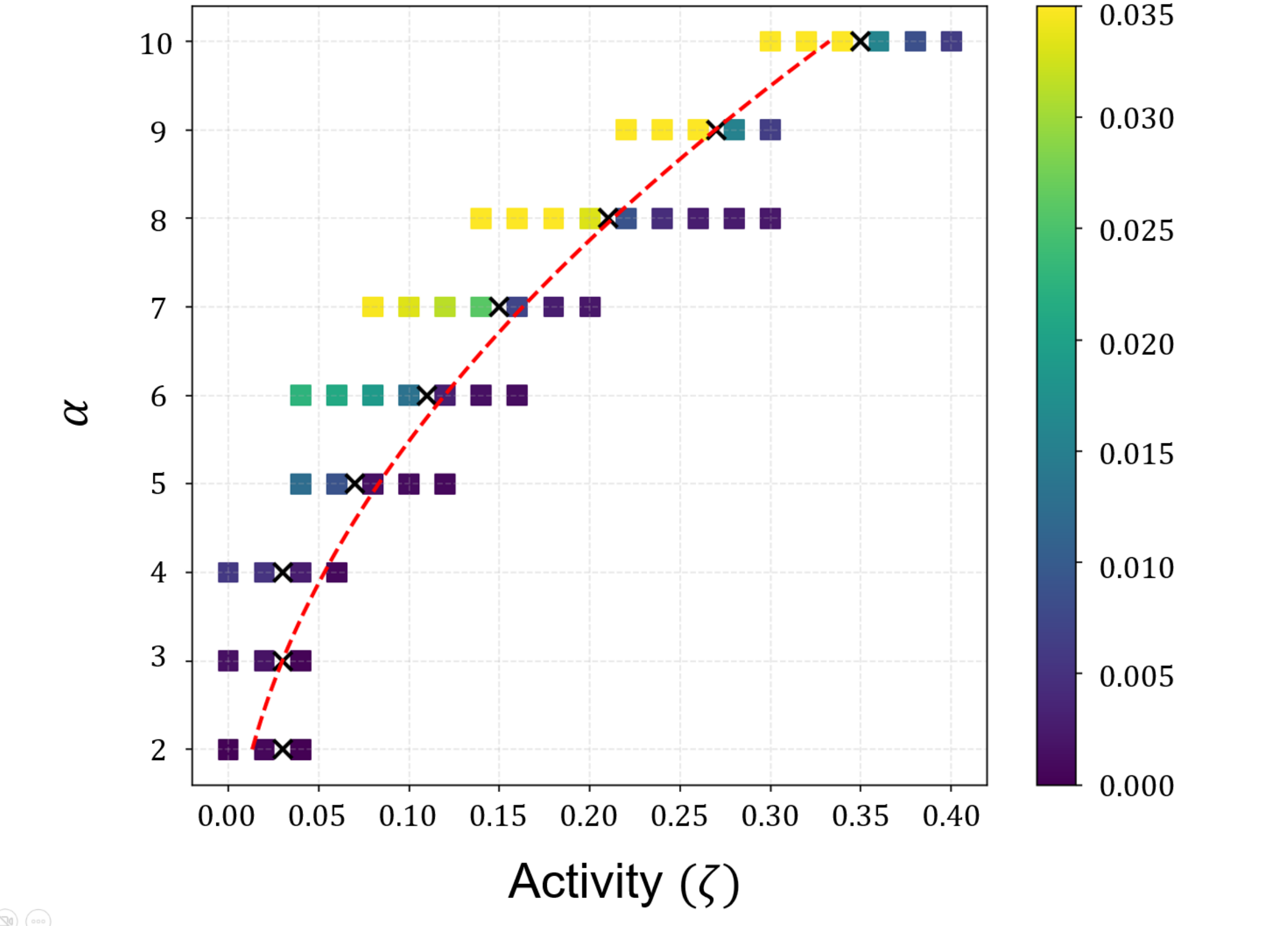}
    \caption{Phase diagram showing the time-averaged bending energy $\langle f_{\text{bend}} \rangle$ as a color map in the $(\alpha, \zeta)$ plane. The energy is averaged over the interval $t = 1000$ to $10000$. Crosses ($\times$) indicate transition points, defined as the intermediate activity values where the slope of $\langle f_{\text{bend}} \rangle$ with respect $\zeta$ is steepest These points are consistent within $\Delta\zeta = 0.04$ with the transition thresholds determined from the slope of mean squared velocity $\langle v^2 \rangle$. The red dotted line represents a fit to these transition points, consistent with the scaling $\zeta \propto \alpha^2 / \kappa$.}
    \label{fig4:phase_diagram}
\end{figure}

Fig.~\ref{fig4:phase_diagram} shows the phase diagram in the $(\zeta, \alpha)$ plane. The color scale indicates the spatiotemporally averaged bending energy density
$\langle f_{\text{bend}} \rangle$ 
calculated over the late-stage interval.
There are two distinct regimes. In the low-activity regime (below the critical boundary), the system exhibits high bending energy, 
indicating that the membrane is strongly deformed locally by the defects (the "trapped" state). 
Conversely, in the high-activity regime, the bending energy drops abruptly, suggesting that the membrane becomes 
relatively flat as the defects unbind and move freely. 

The underlying mechanism of this transition can be understood
through a scaling analysis that balances the active driving force against the curvature-induced trapping force. 
Eq.~\eqref{eq:steady_state} implies that the membrane curvature scales as $\nabla^2 h \sim \alpha S_0 / \kappa$, where $S_0$ represents the characteristic magnitude of the nematic order parameter. 
At the transition threshold, the active stress driving the defect ($\sim \zeta S_0$) is balanced by the elastic coupling force 
generated by the membrane 
curvature ($\sim \lambda S_0 \alpha \nabla^2 h$). 
Equating these terms yields a scaling law for the critical activity: $\zeta_c \sim \alpha^2 / \kappa$. 
As shown in Fig.~\ref{fig4:phase_diagram}),
the boundary between the two phases 
is well fitted by this predicted scaling 
$\zeta_c \propto \alpha^2 / \kappa$ (red dashed line), confirming the validity of our scaling analysis. 

\Add{Note that while the onset of instability from a flat and uniformly oriented state ($h=0, \bm{n}=\bm{e}_x$) can be studied via linear stability analysis , both the curvature-orientation coupling~\cite{uchida2002dynamics} and activity~\cite{kinoshita2023flow} independently destabilize this uniform state.
However, the transition observed at $\zeta_c$ in our system occurs between two distinct non-uniform states: a crumpled membrane trapping defect pairs (at low $\zeta$) and an active turbulent state (at high $\zeta$). 
Since developing a linear stability analysis around such complex, non-uniform configurations is analytically intractable, we instead employ a scaling argument to estimate the transition threshold $\zeta_c$.}

\subsection{Interplay between nematic structures and membrane curvature}

\begin{figure*}[t]
    \centering
    \includegraphics[width=\linewidth]{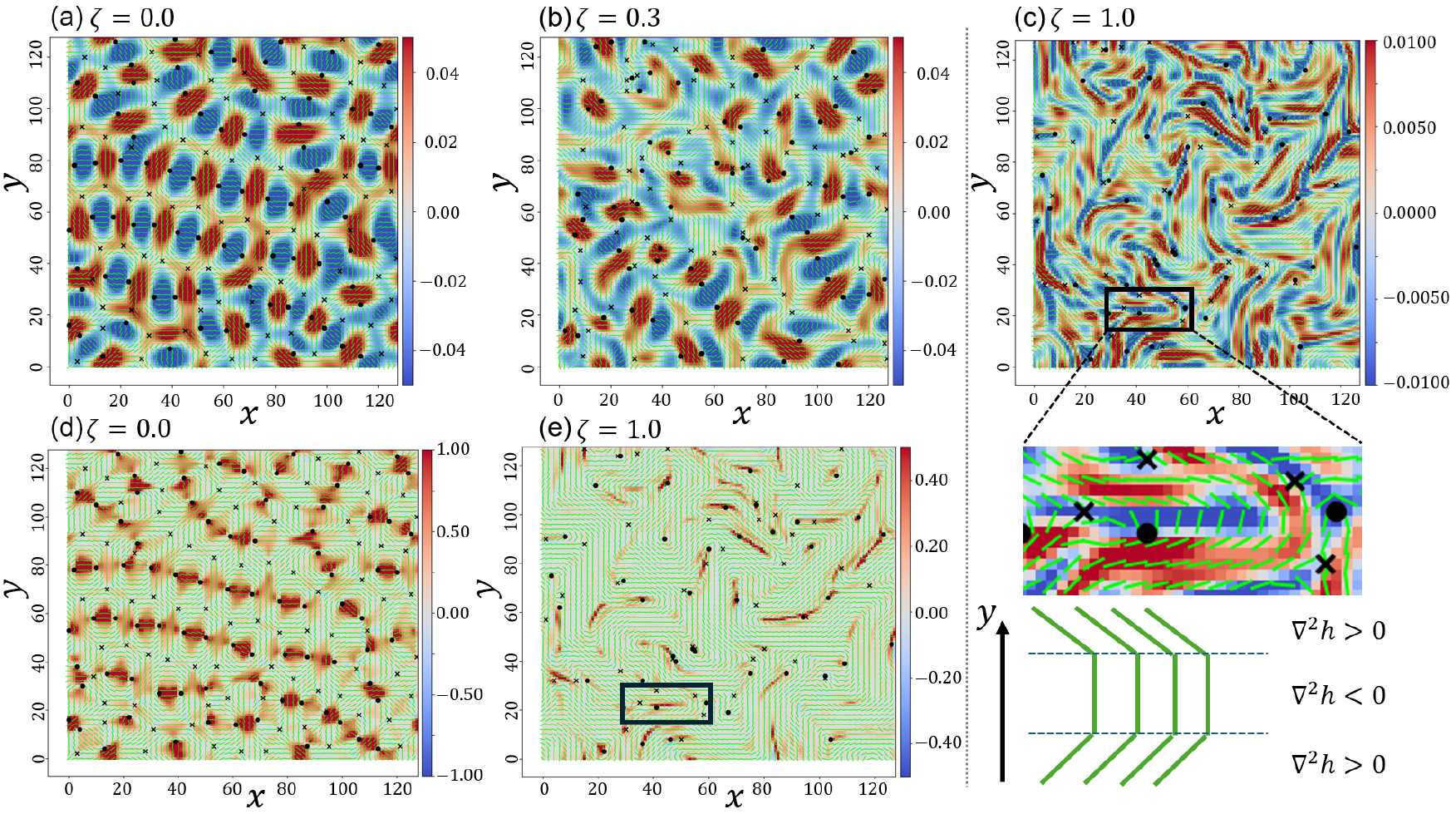}
    \caption{Snapshots of the director field and the mean curvature field for a fixed $\alpha = 10$. Panels (a)--(c) correspond to $\zeta = 0$, $0.3$, and $1.0$, respectively. The background color in (a)--(c) indicates the mean curvature and the green lines represent the director field, while (d) and (e) show the correlation $\nabla^2 h \cdot J(\bm{r})$ between the mean curvature and the source term $J(\mathbf{r}) = \nabla\nabla:\bm{Q}$. Circles ($\circ$) and crosses ($\times$) denote $+1/2$ and $-1/2$ defects, respectively. The region enclosed by the black box highlights an example of the correlation between the director field and the sign of the curvature at the wall.}
    \label{fig5:curvature_pattern}
\end{figure*}

Figure~\ref{fig5:curvature_pattern} displays snapshots of the spatial pattern of the mean curvature $\nabla^2 h$ and its correlation with the nematic texture. 
For vanishing activity (Fig.~5(a)), the observed pattern is 
consistent with the previous results for passive nematic membranes~\cite{uchida2002dynamics}, 
where a $-1/2$ defect is associated with three lobes of positive mean curvature, whereas a $+1/2$ defect exhibits 
a single positive lobe. 
The spatial coincidence between the curvature and the source term $J(r)$ in this passive limit is confirmed in Fig.~5(d), where the product $\nabla^{2}h \cdot J(r)$ is positive throughout the membrane.
Specifically, we identified pairs of defects facing head-to-head in positive curvature regions, as well as tail-to-tail pairs in negative curvature regions. 
Previous analysis~\cite{uchida2002dynamics} shows that a $+1/2$ defect possesses nine times higher bending energy compared 
to a $-1/2$ defect. 
To mitigate this energetic cost, $+1/2$ defects tend to form pairs, thereby reducing the curvature elastic energy by superimposing regions of like-signed source terms in a head-to-head configuration. 
In the head-to-head configuration, the active force drives the defects toward each other, while the Frank elastic repulsion opposes this motion.
Near the critical activity (Fig.~5(b)), although some $+1/2$ defects overcome the curvature-induced trapping potential and become 
self-propelled, the distinct spatial arrangements persist.
The emergence of self-propelled +1/2 defects is explained 
by the active driving force surpassing the elastic binding force,
causing the pairs to dissociate via scattering.

In the high-activity regime (Fig.~5(c)), 
we observe the formation of dynamic, wave-like elongated structures characterized by a central line of negative curvature flanked by parallel lines of positive curvature. These curvature patterns spatially coincide with the "walls" in the director field, which are regions of sharp orientational kinks~\cite{thampi2014instabilities}. 
The strong dynamical coupling between these features is clearly visualized in Fig. 5(e), where the product of the curvature and the source term, $\nabla^2 h \cdot J(r)$, is highly localized and positive along the walls. This spatial correlation persists because the wall structures generate an extremely large local stress that forcibly drives the membrane deformation. 

To gain deeper insight into the mechanism governing these observations, we analyze the force generated by a $+1/2$ defect to deform the membrane. We use the complex form $\mathcal{Q} = Q_{xx} + i Q_{xy} = 
\frac12 S e^{2i \theta}$
and the polar coordinates $(r, \phi)$, 
where $r$ is the distance from the core. We assume that the scalar order parameter depends only 
on the distance from the core, $S=S(r)$. 
By noting that $J = \partial_i \partial_j Q_{ij} = \text{Re} ((\nabla^*)^2 \mathcal{Q})$ and setting $\theta = \phi / 2$ for a $+1/2$ defect, we obtain:
\begin{equation}
    J(\bm{r}) = \frac12 \cos \phi \left( S_{rr} + \frac{1}{r} S_r - \frac{1}{r^2} S \right). 
\end{equation}
Outside the core where $S$ is assumed to be constant,
the expression is simplified to $J \approx -S \cos \phi / (2r^2)$. 
Thus, the +1/2 defect possesses a source term that is negative at the tail ($| \phi | < \pi / 2$) and positive at the head ($| \phi | > \pi / 2$).    
As a result, aligning two +1/2 defects head-to-head reinforces their respective source terms, lowering the curvature elastic 
energy $F_{\text{curv}}$ according to Eq.~\eqref{eq:e_curv}.
The observation that +1/2 defects are located closer to each other near the critical point than in the passive case 
(comparing Figs. 5(a) and 5(b))
is attributed to the action of active stress. 
This phenomenon is analogous to the mechanism reported in~\cite{vafa2022active}, where activity stabilizes a $+1$ bound state (a pair of +1/2 defects).

To examine the relationship between curvature and wall structure, we consider a simplified one-dimensional geometry. 
We define a local coordinate system where the $x$ and $y$ axes correspond to the tangential and normal directions of the wall, respectively. In this setup, all fields vary only along 
the $y$-direction. Let $K(y) = \partial_y^2 h$ denote the local curvature.
In this frame, Eq.~\eqref{eq:steady_state} simplifies to 
$\partial_y^2 K(y) = - (\alpha/\kappa) \partial_y^2 Q_{yy}$.
Integrating twice yields the general solution: 
$K(y) = K_0 \cos 2\theta(y) + c$, where $K_0 \equiv \alpha S/(2\kappa)$ 
and $c$ is an integration constant. 
Requiring the membrane to be flat at infinity
 ($K \to 0$ for $\theta = 0, \pi$) fixes the constant as $c = -K_0$,
 leading to  the analytical expression:
\begin{equation}
K(y) = K_0 (\cos 2 \theta(y) - 1) \leq 0,
\end{equation}
which demonstrates that the curvature is intrinsically negative within the wall structure for $\alpha > 0$. In our periodic simulation, the condition $\langle K(y) \rangle = 0$ modifies the constant to 
$c = - K_0 \langle \cos 2\theta(y) \rangle$. Consequently, 
positive curvature arises outside the wall to compensate for the 
negative curvature within the wall.

Although the analysis presented above is based on a static system, 
this physical picture effectively describes the dynamic scenarios
observed in our simulations for the following two reasons. 
First, the director walls exhibit high temporal stability, persisting for a significant duration relative to the defect motion. This persistence explains why the correlation between the wall and the curvature is observed even in the high-activity regime beyond the critical point, where the membrane deformation typically fails to keep up with the rapid active flow. Furthermore, since the spatial variation of $\bm{Q}$ is abrupt within the wall , the source term $\nabla \nabla : \bm{Q}$ becomes locally enhanced, as visualized in Fig.~\ref{fig5:curvature_pattern}(e). This demonstrates that the wall generates a strong local stress sufficient to forcibly drive the deformation, thereby compensating for the slow relaxation speed of the membrane. Such localized driving allows the membrane shape to remain coupled to the nematic texture even in the turbulent regime, where the slow relaxation of the membrane typically fails to follow the rapid motion of individual defects.

\section{Discussion and Summary}

In this paper, we have investigated the dynamics of active nematic membranes governed by anisotropic curvature coupling. 
Using a minimal model, we identified a continuous transition determined by the competition between the active stress and the curvature-induced potential. 
We derived a scaling law for the critical activity, $\zeta_c \sim \alpha^2/\kappa$, which successfully captures the phase boundary between the defect-trapped regime and the active turbulent regime. The defects trapped by local curvature are released at $\zeta = \zeta_c$; near the \Add{onset of the instability}, the defect velocity $v_d$ is proportional to the net driving force $\zeta - \zeta_c$, which determines the fluid velocity. \Add{This phenomenology of defect trapping at low activity and subsequent release above a threshold shares a striking analogy with active nematics in other structured environments. For instance, microfabricated pillars or elastic inclusions have been shown to act as geometric constraints that create potential wells for topological defects, with activity eventually driving their release~\cite{velez2024probing, santos2025spontaneous}. Our results suggest that membrane curvature plays a functionally equivalent role to such rigid obstacles, providing an energetic mechanism for defect organization and controlled release.}
At higher activity, the mean square of the fluid velocity $\langle v^2 \rangle$ increases linearly with $\zeta$, extending the  scaling law $\langle v^2 \rangle \propto \zeta$ for conventional active nematics~\cite{giomi2015geometry, shankar2019hydrodynamics, hemingway2016correlation}
to systems with deformable geometries.

Crucially, we demonstrated that even in the highly dynamic turbulent regime, the membrane shape remains strongly coupled to the nematic order. 
We revealed that the "walls"~\cite{thampi2014instabilities} 
in the orientational field act as localized sources of stress, driving wave-like deformations of the membrane. 
This mechanism was further examined through an analytical solution of a one-dimensional static model, which qualitatively explains the observed curvature profiles by considering the interplay between local coupling and global geometric constraints. 

Despite these results, our current model relies on several simplifications. First, the use of the Monge gauge restricts the applicability to small-gradient configurations, precluding the description of large deformations or folding. Second, while we considered a free membrane, experimental realizations often involve substrates, necessitating the inclusion of substrate friction or adhesion effects in future studies. Finally, regarding biological relevance, actual organisms like {\it Hydra} possess a bilayer structure~\cite{maroudas2021topological}. Therefore, to fully capture the complexity of {\it Hydra} morphogenesis, 
extensions of the model should account for the mechanical 
interplay between coupled layers.

Nevertheless, our findings highlight the role of topological defects and nematic textures not merely as passive markers of disorder, but as active organizers of geometry. 
This physical framework provides insights into biological processes where defects correlate with morphological features, such as in {\it Hydra} regeneration and epithelial tissue mechanics, suggesting that anisotropic curvature coupling is a key ingredient in defect-mediated 
morphogenesis.

\Add{\section*{Data and Code Availability}
The source code for the numerical simulations and the data supporting the findings of this study are publicly available in Zenodo~\cite{zenodo_data}.}

\bibliography{hirota2026defect}

\end{document}